\documentclass[superscriptaddress,reprint,aps,prl]{revtex4-1}

\usepackage{amssymb, amsmath}
\usepackage{graphicx}
\usepackage{dcolumn}
\usepackage{bm}

\begin{document}

\preprint{APS/PRL}

\title{Spin-orbital excitation continuum and anomalous electron-phonon interaction in the Mott insulator LaTiO$_3$}

\author{C.~Ulrich}
\affiliation{School of Physics, University of New South Wales, Sydney, NSW 2052, Australia}
\affiliation{Max-Planck-Institut~f\"{u}r~Festk\"{o}rperforschung, Heisenbergstr.~1, D-70569 Stuttgart, Germany}
\author{G.~Khaliullin}
\affiliation{Max-Planck-Institut~f\"{u}r~Festk\"{o}rperforschung, Heisenbergstr.~1, D-70569 Stuttgart, Germany}
\author{M. Guennou}
\affiliation{Max-Planck-Institut~f\"{u}r~Festk\"{o}rperforschung, Heisenbergstr.~1, D-70569 Stuttgart, Germany}
\affiliation{Materials Research and Technology Department, Luxembourg Institute of Science and Technology, 41 rue du Brill, L-4422 Belvaux, Luxembourg}
\author{H. Roth}
\affiliation{II. Physikalisches Institut, Universit\"at zu K\"oln, 50937 K\"oln, Germany}
\author{T. Lorenz}
\affiliation{II. Physikalisches Institut, Universit\"at zu K\"oln, 50937 K\"oln, Germany}
\author{B.~Keimer}
\affiliation{Max-Planck-Institut~f\"{u}r~Festk\"{o}rperforschung, Heisenbergstr.~1, D-70569 Stuttgart, Germany}

\date{\today}

\begin{abstract}
Raman scattering experiments on stoichiometric, Mott-insulating LaTiO$_3$ over a wide range of excitation energies reveal a broad electronic continuum which is featureless in the paramagnetic state, but develops a gap of $\sim 800$\,cm$^{-1}$ upon cooling below the N\'eel temperature $T_N = 146$ K. In the antiferromagnetic state, the spectral weight below the gap is transferred to well-defined spectral features due to spin and orbital excitations. Low-energy phonons exhibit pronounced Fano anomalies indicative of strong interaction with the electron system for $T > T_N$, but become sharp and symmetric for $T < T_N$. The electronic continuum and the marked renormalization of the phonon lifetime by the onset of magnetic order are highly unusual for Mott insulators and indicate liquid-like correlations between spins and orbitals.
\end{abstract}

\pacs{75.25.Dk, 63.20.kd, 75.30.Et, 71.70.Ch}

\maketitle
Frustrated exchange interactions in magnetic insulators can give rise to ``quantum liquid'' ground states with excitation continua fundamentally different from the dispersive collective modes in magnetically ordered states \cite{Balents}. A lot of recent attention has focused on systems with both spin and orbital degeneracy, where orbital- and bond-selective exchange interactions greatly enhance the degree of frustration \cite{Khal05}. In many potential model materials, the orbital degeneracy is lifted by distortions of the crystal lattice, which can stabilize orbitally and magnetically ordered ground states and thus pre-empt the formation of genuine ``spin-orbital liquids''. Nonetheless, the phase space for spin-orbital fluctuations remains large if the electron-lattice interaction is weak, and these fluctuations can profoundly influence the macroscopic properties. The experimental characterization and theoretical description of such highly degenerate quantum many-body systems is currently at the forefront of research in condensed matter physics \cite{Balents,Khal05}.

Mott-insulating titanates with $3d^1$ electron configuration and pseudocubic perovskite structure have served as model systems for the exploration of spin-orbital quantum fluctuations \cite{Khal05}. The valence electron resides in the threefold degenerate $t_{2g}$ orbital manifold of the nearly cubic crystal field. Crystallographic distortions partially lift this orbital degeneracy \cite{Pavarini}, but the orbital excitation profiles determined by Raman \cite{Ulri06} and resonant inelastic x-ray scattering \cite{Ulri08} are much broader than conventional crystal-field excitations. The question whether this broadening arises from coupling to spin excitations \cite{Ament,Wohlfeld} or phonons \cite{Schmidt,Ishihara} has not been resolved. The reduced ordered moment and isotropic spin wave dispersions in the magnetically ordered states of antiferromagnetic LaTiO$_3$ \cite{Keim00} and its ferromagnetic counterpart YTiO$_3$ \cite{Ulri02} provide indirect evidence of low-energy spin-orbital fluctuations. Thermal conductivity data have been interpreted as evidence of an orbital-liquid state in the paramagnetic phase \cite{Goodenough}, and the thermal expansion provides evidence of coupling between low-energy orbital excitations and the crystal lattice \cite{Knafo}. As spectral information about these excitations has not been available, however, quantitative models of the coupling mechanism have not been proposed. More conventional models based on separate spin and orbital sectors have also been proposed to explain the physical properties of the Mott-insulating titanates \cite{Fritsch,Haverkort}.

Here we report a comprehensive, high-resolution Raman scattering study of lattice vibrations and their interaction with electronic excitations in single crystals of Mott-insulating LaTiO$_3$. In the paramagnetic state, the phonon profiles exhibit pronounced Fano lineshape asymmetries indicative of coupling to a continuum of electronic excitations with energies well below the Mott-Hubbard gap. In the antiferromagnetic state, a gap opens up in the electronic continuum, and the phonon anomalies disappear. The temperature evolution of the electronic continuum and the phonon linewidths indicate liquid-like correlations between spins and orbitals \cite{Khal05,Khal00}.

The polarized Raman experiments were performed on a high quality single crystal of LaTiO$_3$ grown by the floating zone technique \cite{Cwik03}. Its N\'eel temperature, $T_N = 146$ K, is very close to the highest value reported in the literature, indicating a fully stoichiometric composition.
LaTiO$_3$ crystallizes in the orthorhombic GdFeO$_3$ structure (space group $Pbnm$) \cite{Suppl}. Our crystal is twinned within the $ab$-plane due to the small difference between the lattice parameters of the $a$- and $b$-axes, but untwinned along the $c$-axis. The 514.5\,nm line of an Ar$^+$/Kr$^+$ mixed-gas laser was used for excitation, and the scattered light was analyzed using a Dilor-XY triple spectrometer in backscattering geometry. The spectrometer resolution (2.6\,cm$^{-1}$) was determined by a calibration measurement and taken into account through deconvolution
to determine the intrinsic phonon linewidths reported below. To avoid heating of the sample surface, the power of the incident laser beam was kept below 10\,mW with a spot size of diameter 100\,$\mu$m at the sample position.
The measurements presented here were performed with the wavevector of the incident light parallel to the $b$-direction in the $zz$ polarization geometry, that is, both incident and scattered photons were polarized along the $c$-axis.

\begin{figure}[b]
\centering
\includegraphics[width=8.2cm]{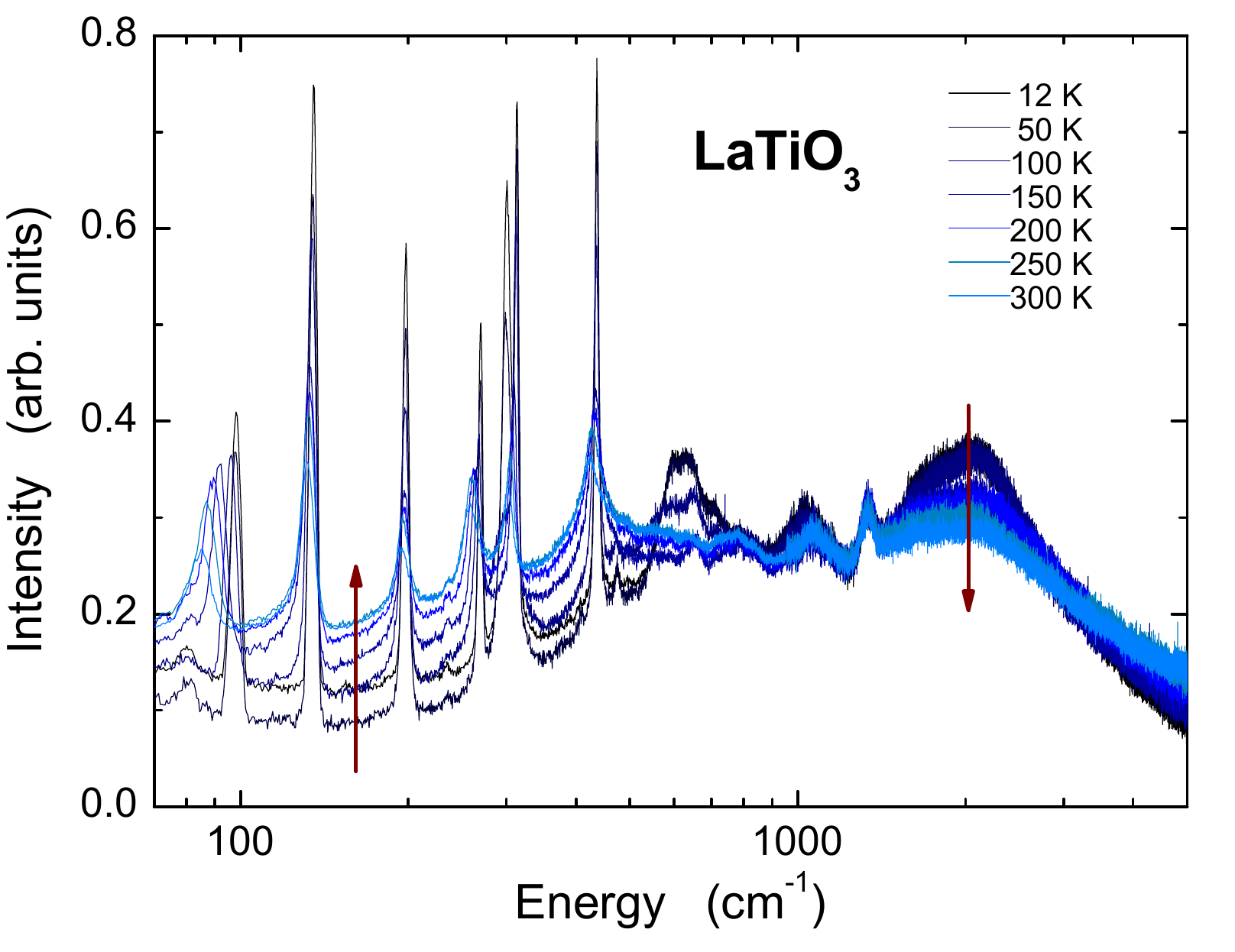}
\caption{
(color online)
Raman spectrum of LaTiO$_3$ at various temperatures measured in $zz$ polarization. Note the logarithmic scale of the energy axis.}
\end{figure}

\begin{figure}[t]
\centering
\includegraphics[width=8.2cm]{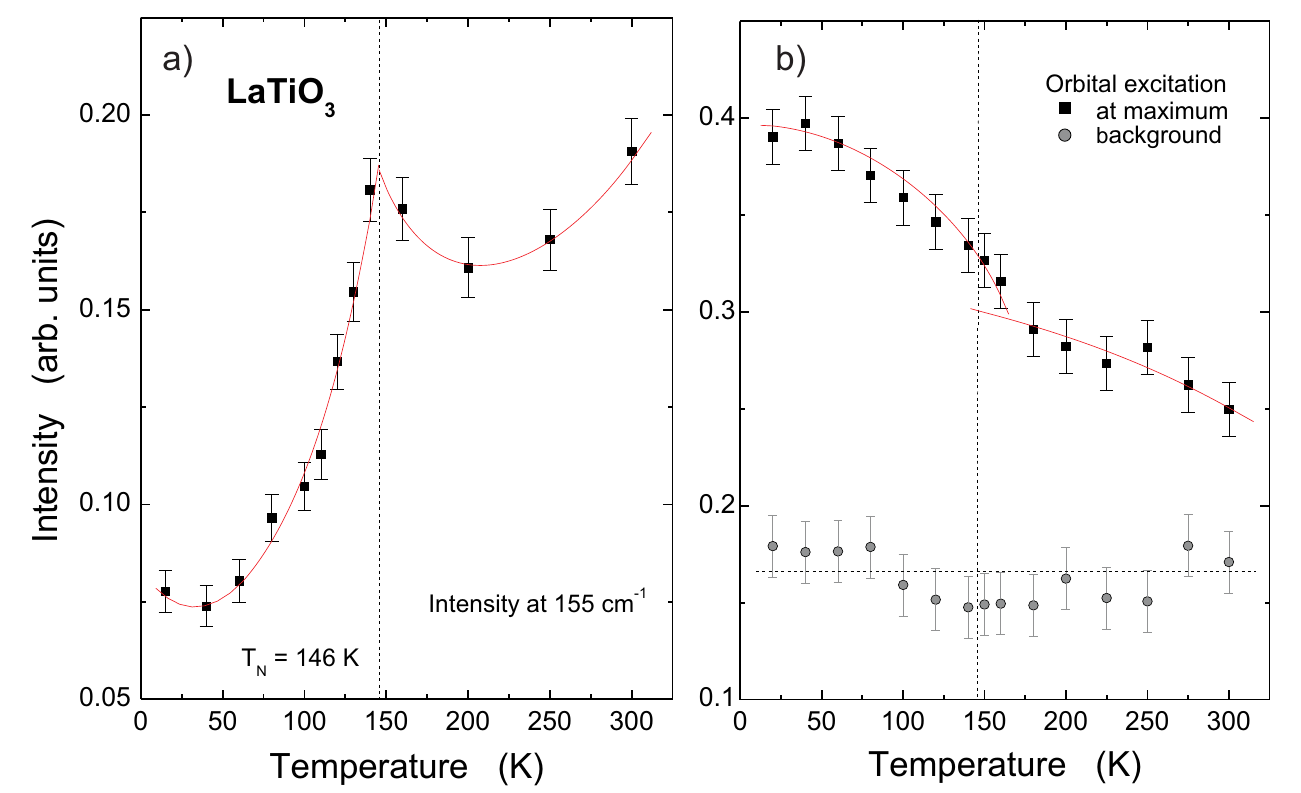}
\caption{
(color online)
Temperature dependence of the intensity of the electronic Raman signal recorded (a) in the phonon range at 155\,cm$^{-1}$, and (b) at the orbital excitation energy
2000\,cm$^{-1}$. The background was determined at
4000\,cm$^{-1}$. Pronounced changes are observed for both energies at $T_N$ (vertical dashed lines). The red solid lines are guides-to-the-eye.}
\end{figure}

Figure 1 shows the temperature evolution of the $zz$-polarized Raman spectrum of LaTiO$_3$ over a wide range of photon energies. We first focus on the broad continuum, which extends over a spectral range of at least 50-4000 cm$^{-1}$. Electronic scattering of this kind is highly unusual for insulating solids. A similar continuum was observed in previous Raman scattering experiments on other titanates over a narrower spectral range \cite{Reed97}, but the earlier work was limited to room temperature. Upon cooling below $T_N$, our data reveal the formation of a gap-like structure for energies below $\sim 800$ cm$^{-1}$. The strong temperature dependence confirms the intrinsic origin of the continuum, and the marked intensity anomaly at $T_N$ (Fig. 2a) demonstrates that it at least partially originates from spin excitations.

The spectral weight lost below the gap in the antiferromagnetic state accumulates in two well-defined features around 660 and 2000\,cm$^{-1}$. Based on a comparison with neutron scattering data \cite{Keim00}, the former feature can be attributed to two-magnon Raman scattering \cite{Iliev04}. This underscores the association of the electronic continuum in the paramagnetic state with spin excitations. The broad peak around 2000 cm$^{-1}$, on the other hand, which draws an even larger fraction of the continuum spectral weight, is well above the two-magnon range and has been assigned to orbital excitations \cite{Ulri06}. Its electronic origin was confirmed by resonant inelastic x-ray scattering \cite{Ulri08}. Although a rigorous sum rule does not apply in Raman scattering, the marked intensity enhancement of this excitation below $T_N$ (Fig. 2b), which is fed in part from spectral weight below the gap at 800 cm$^{-1}$ (Fig. 2a), suggests mixing of spin and orbital excitations in the electronic continuum.

The lattice vibrations and their temperature evolution provide further insight into the nature of the electronic continuum. Figure 3a provides an overview over the low-energy range of the Raman spectrum, along with the phonon mode assignment \cite{Suppl}. The low-energy phonons exhibit striking lineshape anomalies at $T_N$, which are highlighted in more detail in Fig. 3b for three representative modes. Above $T_N$, the phonon peaks are highly asymmetric and are well described by Fano profiles (lines in Fig. 3b) which indicate strong interaction with the electronic continuum. Upon cooling below $T_N$, the Fano parameters extracted from fits to the experimental data decrease continuously (Fig. 3c), in a manner that parallels the temperature dependence of the magnetic order parameter \cite{Keim00}. As $T \rightarrow 0$, the phonon profiles become sharp and symmetric (Figs. 3b,c).

Figure 4 displays the energies and intrinsic full-widths-at-half-maximum (FWHM) of selected phonons resulting from the fits as a function of temperature. For $T > T_N$, the phonons harden upon cooling and their linewidths decrease, as expected from anharmonic phonon-phonon interactions. Fits to a standard anharmonic-decay model \cite{Klemens} provide excellent descriptions of these data (lines in Fig. 4a-f). Below $T_N$, however, the linewidths of all phonon modes decrease markedly, consistent with the opening of the electronic gap (and consequent loss of electron-phonon decay channels) discussed above, and with the behavior of the Fano parameter (Fig. 3c). The phonon energies also exhibit anomalies at $T_N$, in qualitative agreement with prior infrared spectroscopy work \cite{Hemb03}. It is interesting to note that all phonon modes with energies below 300 cm$^{-1}$ show a pronounced change in their asymmetry and shift to higher energies below $T_N$, whereas phonon modes at higher energies soften and remain symmetric in the entire temperature range. The crossover between both behaviors coincides with a sharp mode with energy $\sim 300$ cm$^{-1}$ that is only present below $T_N$ (Figs. 3a and 4e,f). Based on a comparison with neutron scattering data \cite{Keim00}, this mode can be tentatively identified with a single-magnon excitation at the boundary of the pseudocubic unit cell, which is folded to the Brillouin zone center in the orthorhombic space group $Pbnm$ and hence gives rise to Raman-active magnon excitations. The question whether or not the coincidence of these two energies is accidental remains an interesting subject of future investigation.

We now discuss the origin of the unusual behavior of the phonon profiles. Fano lineshapes are commonly encountered in Raman spectra of metals, and phonon lineshape anomalies of similar amplitude have been observed for spin density wave transitions in correlated metals such as the iron arsenides \cite{Rahlenbeck}. LaTiO$_3$, however, is an insulator with a Mott-Hubbard gap of $\sim 200$ meV \cite{Okim95}, well above the phonon range. Optical absorption experiments have uncovered a weak tail of charge excitations (possibly arising from localized states in the gap) that extends down to 50 meV (400 cm$^{-1}$) \cite{Lunkenheimer}, but this is still well above the energies of the phonons with the most pronounced anomalies (Fig. 2). Moreover, the onset energy of these weak optical excitations does not change significantly across $T_N$ \cite{Lunkenheimer}, in contrast to the dramatic renormalization of the phonon lineshapes shown in Figs. 2-4. Charge excitations thus cannot be responsible for the phonon anomalies reported here.

Another possible origin of the phonon anomalies is coupling to spin excitations. Magnetic-order induced phonon frequency shifts of the order of $\delta \omega / \omega \sim $ 1~\%
are indeed commonly observed in Mott insulators. Two channels contribute to the magnetoelastic coupling. The first one operates via an unquenched part of the orbital angular momentum
that links the magnetic order parameter to the lattice. This mechanism is effective for magnetic ions with
large spin-orbit coupling, but is unlikely to dominate in titanates where the spin is only
one-half (no single-ion anisotropy), the spin-orbit coupling is weak, and
the orbital angular momentum is quenched \cite{Keim00}. Alternatively, two-ion contributions
to the magnetoelastic coupling might be important. This channel operates via the modulation of the spin
exchange energy $H_s = J_s \, {\vec S}_i{\vec S}_j $ by lattice displacements $u$:
$J_s= J_s^{(0)} + J^{\prime}_s \, u + ... \ \ \ $.
Spin-order induced phonon shifts have been observed in Mott-insulating LaMnO$_3$ \cite{Gran99,Quij01}
and were discussed within this framework. The phonon shifts observed here can be understood in a similar way.

\begin{figure}
\centering
\includegraphics[width=8.2cm]{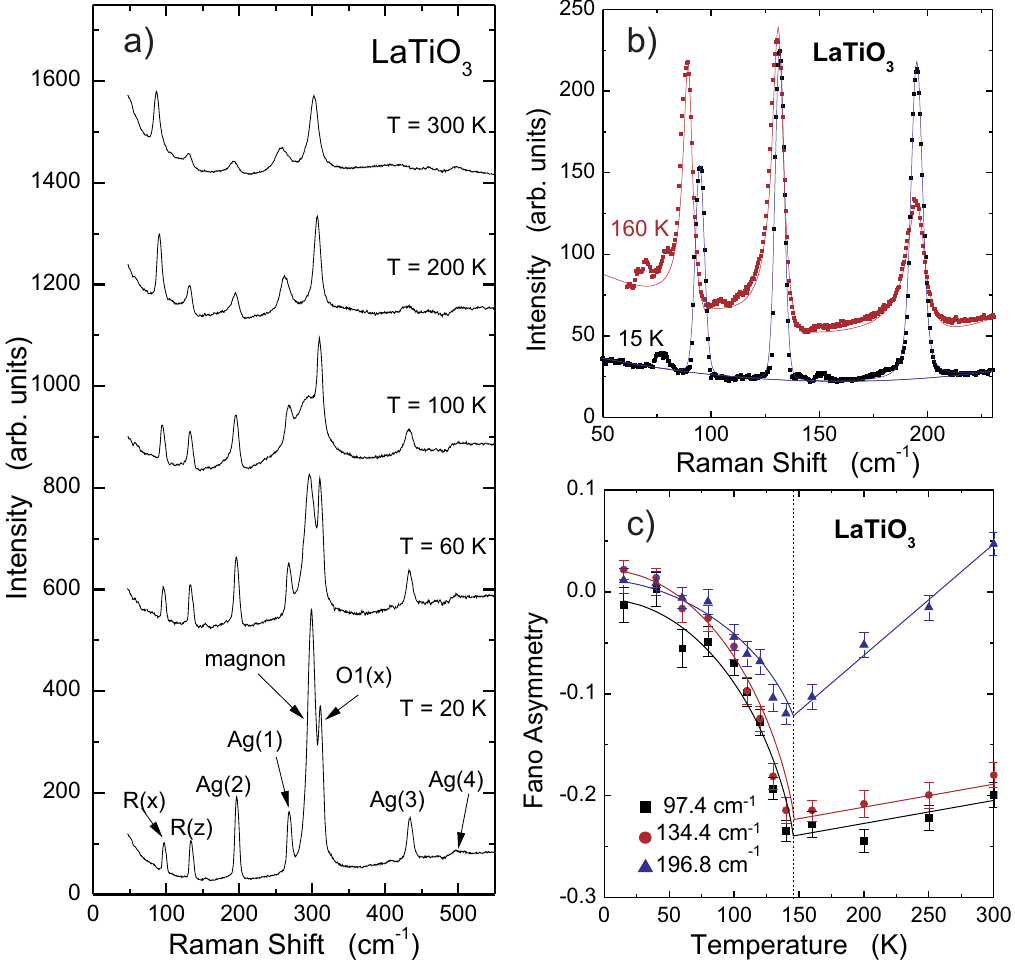}
\caption{
(color online)
(a) Low energy phonon part of the Raman spectrum of LaTiO$_3$ at various temperatures. See the Supplemental Material for the mode assignment \cite{Suppl}. (b) Low-energy phonons above and below $T_N$. The lines are the results of fits to Fano profiles. (c) Temperature dependence of the Fano asymmetry parameter. The vertical line indicates $T_N$. The solid lines are guides-to-the-eye.}
\end{figure}

\begin{figure}
\centering
\includegraphics[width=8.2cm]{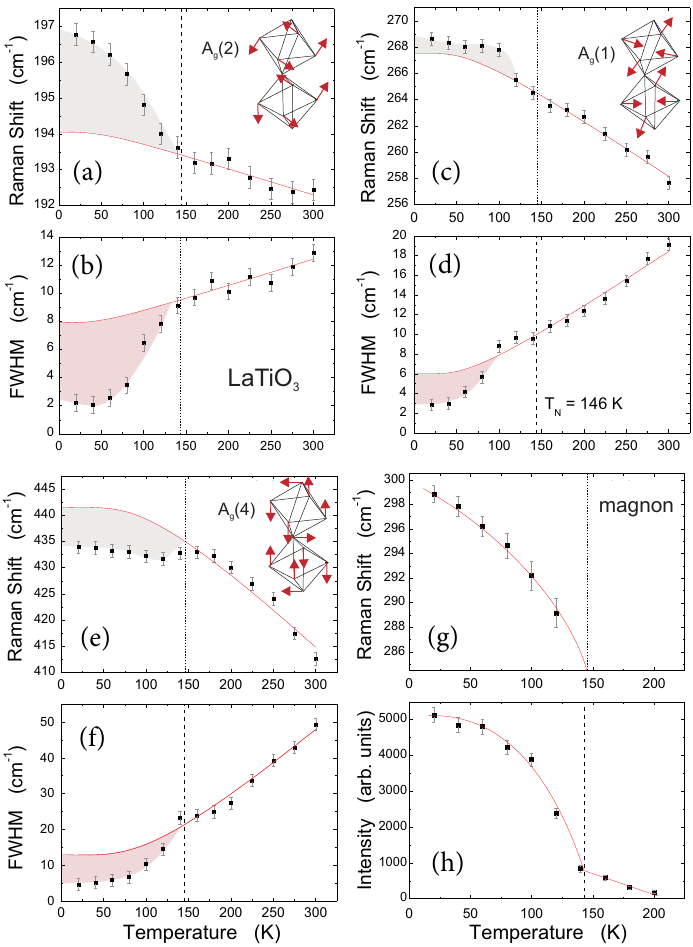}
\caption{
(color online)
Temperature dependence of (a, c, e) peak energy and (b, d, f) linewidth (FWHM) of selected $A_g$ phonon modes and (g, h) the first-order magnon
at ~300\,cm$^{-1}$ (37.2\,meV). The red lines correspond to the temperature dependence in case of anharmonic
phonon-phonon interactions. The linewidths of all observed phonons decrease below $T_N$ (vertical dashed lines), and the phonon energies deviate strongly from the behavior expected from pure phonon-phonon interactions below this temperature. The red solid lines for the magnon serve as guides to the eye.}
\end{figure}

However, the drastic shortening of the phonon lifetime with spin-order melting observed in LaTiO$_3$ is highly unusual.
In the extensively studied manganites, for instance, the phonon linewidth does not show any anomalies at the magnetic transition \cite{Gran99,Quij01},
and we are not aware of such a large ($\delta \Gamma/ \omega \sim $ 3~\%) change
in the phonon linewidth $\Gamma$ induced by spin disorder in any other Mott insulator with $3d$ valence electrons.
The relative insensitivity of the phonon linewidth to spin ordering is due to the smallness of
the dynamical spin-phonon coupling constant.
The spin-only contribution to the phonon linewidth (FWHM) can be estimated as follows:
$\Gamma / \omega \simeq ( z^{\prime} J^{\prime \, 2}_s   / K) \, \chi_{s}^{\prime\prime}(\omega)$,
where $K$ stands for the force constant relevant for the phonon of interest,
and $z^{\prime}$ counts the number of spin-bonds affected by the vibration pattern ({\it e.g.},
the octahedral tilts/rotations that modulate the Ti-O-Ti bonds). The phonon linewidth is then
determined by the spectral density $\chi_{s}^{\prime\prime}(\omega)$ of the
operator $\hat{B}_s={\vec S}_i{\vec S}_{i+\delta}$, which describes spin fluctuations at this phonon
frequency. Assuming a featureless spin fluctuation
spectrum in the paramagnetic state, this can be roughly estimated as $\chi_{s}^{\prime\prime} \sim \omega (S/zJ_s)^2$,
with $z$ being the coordination number. Considering $\omega \leq zJ_s$ and $z^{\prime} \simeq z$,
this results in  $\Gamma/\omega \sim (J^{\prime}_s S)^2/J_sK = \alpha^2 (J_sS^2)/K$, where the
parameter $\alpha$ was introduced as $J^{\prime}_s = \alpha J_s /$\AA. With $J_s \sim 16$~meV \cite{Keim00} and typical values
$K \sim 30$~eV$/$\AA$^2$, we find $\Gamma/\omega \sim 10^{-4}\alpha^2$
which, even for values as large as $\alpha = 5$, gives only a $\sim$ 0.3~\% effect.
This explains why minimal changes in $\Gamma$ are seen in the manganites below $T_N$ \cite{Gran99,Quij01}. The
extraordinarily large variation of $\Gamma$ induced by spin disorder in
LaTiO$_3$ is thus difficult to explain in terms of spin-only models.

Having excluded charge and spin fluctuations, orbital excitations remain as the only viable origin of the phonon lifetime anomalies. Such excitations couple directly to phonons via modulation of the chemical bonds, and this dynamical coupling is in general stronger than
the rather indirect spin-lattice coupling discussed above.
Concerning the orbital excitations in titanates, two opposite views have been proposed:
({\it i}) the orbital levels are split by lattice distortions, eliminating orbital fluctuations
for energies $\leq 200$~meV \cite{Hemb03,Moch03};
({\it ii}) the orbitals are more closely confined to the spins (via the superexchange process)
than to the lattice, and their fluctuations extend to low energies including the phonon
range \cite{Keim00,Khal00}. The spin-order-selective phonon lifetime anomalies reported here clearly favor the latter scenario.

In the theory developed in Refs. \cite{Khal05,Khal00}, the spin-only Hamiltonian
$J_s \, {\vec S}_i{\vec S}_j$ is augmented by simultaneous permutations of both spin and
orbital quantum numbers (the latter represented by bond-dependent pseudo-spin
operators ${\vec \tau}_i$), which are promoted by virtual particle-hole
charge excitation across the optical gap. Because of the larger number of states (two spin {\it times}
three orbital), and due to strong
frustration inherent to orbital interactions, quantum effects are enhanced, and
the ground state is dominated by spin-orbital bond fluctuations analogous to the resonating-valence-bond picture \cite{Ande87}.
Perturbations such as the Hund coupling induce long-range spin order (and possibly orbital order \cite{Goodenough}) below $T_N$, and a gap opens up in the spin-orbital excitation spectrum. Nonetheless, the high-energy/temperature physics is still dominated by the quantum motion of electrons
in which the spins and orbitals are confined to each other.

Phonons are scattered by these composite excitations either through direct
modulation of the exchange-bonds via the interaction Hamiltonian $J^{\prime} \hat{B}_{so} u$
where $\hat{B}_{so}=2 \big({\vec S}_i\cdot{\vec S}_j+\frac{1}{4}\big)
\big(\vec \tau_i\cdot\vec \tau_j+\frac{1}{4}n_i n_j\big)$, or via the
orbital-lattice coupling, which in simplest form can be written as $g \, Q \, u$ where $Q$
denotes the most relevant component (for a given phonon mode) of the orbital quadrupole moment
tensor. In analogy to the considerations above, one can write for the phonon linewidth in the latter case:
$\Gamma / \omega = (g^2 / K) \, \chi_{Q}^{\prime\prime}(\omega)$ where $\chi_{Q}^{\prime\prime}(\omega)$ is the spectral density of the orbital fluctuations, with the gap opening below $T_N$. Since the direct orbital-phonon coupling constant, $g$, is generically larger than the exchange-modulation parameter $J^{\prime}$, the short electron-phonon lifetimes and their spin-sensitivity are naturally explained in this scenario.

Interestingly, phonon anomalies of similar strength
have been observed in Ca$_2$RuO$_4$, which undergoes successive metal-insulator and antiferromagnetic transitions upon cooling. Whereas the most pronounced anomalies occur at the metal-insulator transition, a renormalization of the phonon lifetime was also observed at the magnetic transition, and was attributed to orbital fluctuations \cite{Rho05}. The mechanism underlying this effect is likely different from the one in LaTiO$_3$. Because of the larger spin-orbit coupling for the $4d$ valence electrons and the larger spin $S=1$, the orbital angular momentum of the
Ru ions is not quenched, and the orbital contribution to the magnetic
order parameter and magnetic fluctuations leads to an enhanced dynamical magnetoelastic
coupling sensitive to the magnetic ordering. The phonon lifetimes in LaTiO$_3$, on the other hand, are determined by {\it intersite}
spin/orbital exchange interactions, rather than the intra-ionic spin-orbit coupling. Nevertheless, the proximity to the metallic state is common to both compounds. It is the large intensity of virtual charge transitions across the optical gap in the titanates that leads to liquid-like
spin/orbital exchange correlations, which are spin-sensitive and ultimately responsible for the observed
phonon anomalies.

We are indebted to M. Cardona and P. Horsch for fruitful discussions. We acknowledge financial support of the DFG under grant TRR80 and the Australian Research Council (ARC) through the Discovery Projects funding scheme (No. DP110105346).


\end{document}